\documentclass[10pt]{article}
\usepackage{amsmath}
\usepackage{color}
\usepackage{amsmath, dsfont}
\usepackage{amssymb}
\usepackage[dvips]{graphics}
\usepackage{epsfig}
\usepackage{calc}
\usepackage{amsfonts}
\usepackage{graphicx}
\usepackage[ansinew]{inputenc}

\title{ Regular black holes via the Kerr-Schild construction\\ in DHOST theories}
\author{Eugeny Babichev $^{1}$, Christos Charmousis $^{1}$, Adolfo Cisterna $^{2,3}$ and Mokhtar Hassaine $^{4}$\\
$^{1}$ Universit\'e Paris-Saclay, CNRS/IN2P3, IJCLab, 91405 Orsay, France \\
$^{2}$ Vicerrector\'ia Acad\'emica, Toesca 1783, Universidad Central
de Chile, Santiago, Chile.\\
$^{3}$ TIFPA - INFN, Via Sommarive 14, 38123 Povo (TN), Italy.\\
$^{4}$ Instituto de Matem\'{a}tica y F\'{\i}sica, Universidad de
Talca, Casilla 747,Talca, Chile.} \textheight 16 cm\textwidth 16 cm
\let\ssection=\section
\renewcommand{\section}{\setcounter{equation}{0}\ssection}

\newcommand{\be}{\begin{equation}}
\newcommand{\ee}{\end{equation}}
\newcommand{\bea}{\begin{eqnarray}}
\newcommand{\eea}{\end{eqnarray}}
\newcommand{\beal}{\begin{aligned}}
\newcommand{\eeal}{\end{aligned}}

\newcommand\B{{\mathcal{B}}}
\newcommand{\HH}{{\mathcal{H}}}
\newcommand{\A}{{\mathcal{A}}}
\newcommand{\ZZ}{{\mathcal{Z}}}

\renewcommand\a{\alpha}

\renewcommand\b{\beta}

\begin{document}
\maketitle
\begin{abstract}
We extend the standard Kerr-Schild solution generating method  to higher order scalar
tensor theories that are shift-invariant for the scalar field. Certain degeneracy conditions, crucial
for the absence of Ostrogradski ghosts, are found to be required for the validity of the Kerr-Schild
ansatz while on the other hand, theories with no parity symmetry are excluded from the solution generating method. The extended
Kerr-Schild symmetry turns out to be a very useful tool to easily
construct black hole solutions from  simple seed configurations.
In particular, the generating method developed is adapted to construct generic black holes but also
regular black hole solutions within Degenerate Higher Order Scalar Tensor (DHOST) theories.
As a particular example we show how to construct explicitly the Hayward metric as a solution to a
specific DHOST theory.
\end{abstract}

\section{Introduction}
Black holes are one of the most intriguing objects in theories of gravitation, and their interest has been considerably
reinforced with the recent detection of gravitational waves
 coming from the coalescence of rotating black holes.
 In this context, a
good understanding of their physical properties  and construction would be crucial
for the comprehension and interpretation of recent and future
observations. This remark applies particularly to the prototype Kerr metric
\cite{Kerr:1963ud} which is known to be the unique stationary and
axisymmetric black hole solution of the vacuum Einstein equations.
It is well-known that the derivation of the Kerr solution is a
highly nontrivial task, due to the
nonlinear character of Einstein's partial differential equations.
One of the ways to face this problem is to consider an ``exact perturbation"
about the flat Minkowski metric $\eta$  along a null direction as
\begin{eqnarray}
g_{\mu\nu}=\eta_{\mu\nu}-2H(x)l_{\mu}l_{\nu},\label{originalKS}
\end{eqnarray}
where $l_{\mu}$ is the tangent vector to a shear-free and geodesic
null congruence and $H$ is a scalar function
\cite{KS,Stephani:2003tm}. This metric ansatz commonly known as the
original Kerr-Schild ansatz presents the advantage of linearizing
the Ricci tensor, and hence the vacuum Einstein equations reduce to
a linear system of equations. Nevertheless,  it is important to note
that only a very specific selection of the shear-free and geodesic
null congruences of the Kerr-Schild ansatz (\ref{originalKS}) will
reproduce the Kerr solution.

The original Kerr-Schild ansatz can also be extended for a not
necessarily flat metric as follows,
\begin{eqnarray}
{g}_{\mu\nu}=g_{\mu\nu}^{(0)}-2 H(x)\, l_{\mu}l_{\nu},
\label{generalkerrschild}
\end{eqnarray}
where $g_{\mu\nu}^{(0)}$ is called the seed metric and $l$ is a null
and geodesic vector field with respect to both metrics, i. e.
$$
{g}^{\mu\nu}l_{\mu}l_{\nu}={g^{(0)}}^{\mu\nu}l_{\mu}l_{\nu}=0,\qquad
(\nabla_{\mu}l_{\nu})l^{\mu}=(\nabla^{(0)}_{\mu}l_{\nu})l^{\mu}=0.
$$
As significative examples of Kerr-Schild metrics, i. e., metrics
that are represented through the Kerr-Schild ansatz
(\ref{generalkerrschild}), we can mention the pp (resp. AdS) wave
background which describes exact gravitational waves propagating in
the Minkowski (resp. AdS) space, with a flat (resp. AdS) seed
metric. As already mentioned, another interesting class of solutions
that may fit within the Kerr-Schild ansatz is provided by static or
spinning black hole solutions. Indeed, it is appealing that in
vacuum, most of the metrics describing black holes are of the
Kerr-Schild form in the sense that there exists a coordinate system
where the metric $g$ can be written as Eq.
(\ref{generalkerrschild}). A remarkable exception is given by the
five-dimensional black ring solution which has a rather distinctive
topology, \cite{Emparan:2001wn}.

In the case of black holes, the Kerr-Schild ansatz
(\ref{generalkerrschild}) can also be seen as a geometrical way of
introducing the mass of the black hole through the scalar function
$H$ while the seed metric corresponds to the asymptotic spacetime
region. The other parameters of the black hole must be encoded in a
nontrivial way in the seed metric. For example, in the case of the
higher-dimensional Kerr-(A)dS metrics \cite{Gibbons:2004uw,
Gibbons:2004js}, the seed metric is described by the (A)dS spacetime
written in ellipsoidal coordinates in the presence of some
parameters $J_i$ whose interpretation as angular velocities will be
effective only for the resulting Kerr-Schild metric. In addition to
providing a geometrical and physical interpretation of the metric,
the Kerr-Schild ansatz has also been proven to be a powerful tool in
deriving stationary black hole solutions of vacuum Einstein
equations with or without cosmological constant \cite{Kerr:1963ud,
KS, Gibbons:2004uw,Gibbons:2004js}. These examples highlight the
importance of the Kerr-Schild ansatz in  vacuum. This is in contrast
to the case  involving
 matter sources, where now the picture becomes rather more complicated. In
fact, the difficulty with the presence of source is to find an
appropriate ansatz for the extra dynamical fields to be fully
compatible with the equations of motion. For example, it is simple
to show that for the Einstein-Maxwell equations, demanding the gauge
potential to be proportional to the null and geodesic vector field
will be consistent only in four dimensions \cite{Newman:1965my,
Debney:1969zz}. In higher dimension, $D>4$, such an ansatz for the
potential field is incompatible with the equations of motion, and
this is clearly one of the reasons behind the lack of an explicit
higher-dimensional Kerr-Newman solution.

In spite of the difficulties of implementing a Kerr-Schild ansatz with
matter, we would like to investigate its feasibility in the somewhat analogous case of scalar tensor theories.  Here the gravitational
dynamical fields are given by the metric $g$  and a scalar field denoted by $\phi$. Scalar tensor
theories are one of the simplest modified gravity theories  which extend General Relativity with one (or more) scalar degrees
of freedom. Quite a few years back, Horndeski presented the most general
(single) scalar-tensor theory with second order equations of motion \cite{Horndeski:1974wa}. The
requirement not to have more than two derivatives in the equations
of motion is connected to an extension of Ostrogradski's no go theorem \cite{Woodard:2006nt}, which states
that, under some assumptions, higher-order derivative theories are unstable acquiring (an Ostrogradski) ghost degree of freedom.
More recently however, it has been shown that some particular
degenerate versions of higher-order scalar-tensor theories can
propagate healthy degrees of freedom. The most general such
Lagrangian depending quadratically on second-order derivatives of a
scalar field was constructed in Refs. \cite{Langlois:2015cwa,
Crisostomi:2016czh}, and dubbed Degenerate Higher Order Scalar
Tensor (DHOST) theory (or extended scalar tensor theory
(EST))\footnote{A classification up to cubic order in the
second-order derivative of the field equations also exists
\cite{Motohashi:2016ftl, BenAchour:2016fzp}.}.
Such DHOST theories have very interesting and rich phenomenology, as
we will also see here. In particular, there exists a subclass of
DHOST theories where gravitational waves propagate at the speed of
light in perfect agreement with the observed results
\cite{Creminelli:2017sry} (see also the critical analysis of
\cite{deRham:2018red}). This DHOST sector has recently attracted a
lot of attention in particular for the search of spherically
symmetric as well as rotating stealth black hole solutions, see
Refs. \cite{{Babichev:2017guv, Minamitsuji:2019tet}} with asymptotic
dark energy properties.

Our guiding example in order to implement the Kerr-Schild procedure within scalar tensor theories
will be the Kerr-Newman solution. In this case, because of the presence of the extra gauge field $A_{\mu}$, we
have to extend the notion of a seed metric to a seed configuration defined as
\begin{eqnarray}
&&ds_0^2=-dt^2+\frac{\Sigma(r,\theta)}{r^2+a^2} dr^2+\Sigma(r,\theta)d\theta^2+(r^2+a^2)\sin^2\theta\, d\varphi^2,\qquad \Sigma(r,\theta)=r^2+a^2\cos^2\theta,\nonumber\\
&&A^{(0)}_{\mu}dx^{\mu}=0.
\end{eqnarray}
Using this seed configuration, the Kerr-Newman solution
in the Kerr-Schild representation (\ref{generalkerrschild}) reads
\begin{eqnarray}
&&ds^2=ds_0^2-\frac{2r}{\Sigma(r,\theta)}\left(q^2-\frac{\mu}{2r}\right)l\otimes l,\qquad l=dt+\frac{\Sigma(r,\theta)}{r^2+a^2}dr-a\sin^2\theta\,d\varphi\nonumber\\
&&A_{\mu}dx^{\mu}=\frac{q\,r}{\Sigma(r,\theta)}l.
\end{eqnarray}
The reason for detailing this example is to observe that, in spite of the changes of the metric and the gauge potential, the norm of this latter
remains unchanged through the Kerr-Schild transformation, i. e.
\begin{eqnarray}
g^{(0)\mu\nu}A^{(0)}_{\mu}A^{(0)}_{\nu}=g^{\mu\nu}A_{\mu}A_{\nu}=0.
\label{normA}
\end{eqnarray}
Now, the idea is to adapt this  example to scalar tensor theories
with the aim of constructing black holes within the Kerr-Schild
transformation. More precisely, we generate nontrivial black hole
solutions from a simple seed configuration. In this paper, for
simplicity, we will not tackle the case of four dimensional
stationary black hole solutions in scalar-tensor theories. Already
in $D=4$ dimensions such stationary metrics, unlike in GR or special cases
\cite{Ruth}, do not even have the circularity property (see for example the
discussion in \cite{VanAelst:2019kku}). In $D=3$, where circularity
is trivially guaranteed, we will find the relevant solution which turns out to
be the BTZ solution \cite{Banados:1992wn} with an effective
cosmological constant.

We consider DHOST theories invariant under the shift symmetry of the
scalar field $\phi\to \phi+\mbox{constant}$, with Lagrangian of the
form ${\cal L}_{\mbox{{\tiny DHOST}}}:={\cal L}(g,\partial g,
\partial^2 g,\partial\phi,\partial^2\phi)$, and as said before we will restrict
ourselves to the static case with a seed configuration given by
\begin{eqnarray}
ds_0^2=-h_0(r)dt^2+\frac{dr^2}{f_0(r)}+r^2\left(d\theta^2+\sin^2\theta\,d\varphi^2\right),\qquad
\phi^{(0)}(t,r):=qt+\psi^{(0)}(r). \label{seedconf}
\end{eqnarray}
Because of the shift-invariance of the theory, we eventually allow
the scalar field to depend linearly on the time coordinate. For a
static metric, the Kerr-Schild transformation
(\ref{generalkerrschild}) from the seed metric  (\ref{seedconf}) can
be more simply re-written as
\begin{eqnarray}
{g}_{\mu\nu}=g_{\mu\nu}^{(0)}+\mu a(r)\, l_{\mu}l_{\nu},\qquad
l=dt-\frac{dr}{\sqrt{f_0(r)\,h_0(r)}}. \label{kerrschild}
\end{eqnarray}
where $\mu$ is a constant parameter proportional to the mass of
the black hole solution, and $a(r)$ is the scalar function to be determined. The resulting metric (\ref{kerrschild}), after a redefinition of the time coordinate
$$
dt\to dt-\mu a(r)\frac{dr}{\sqrt{f_0(r)h_0(r)}\left(h_0(r)+\mu
a(r)\right)}
$$
acquires a diagonal form
\begin{eqnarray}
ds^2=-\left(h_0(r)-\mu
a(r)\right)dt^2+\frac{h_0(r)\,dr^2}{f_0(r)\left(h_0(r)-\mu
a(r)\right)}+r^2\left(d\theta^2+\sin^2\theta\,d\varphi^2\right),
\label{metrKS}
\end{eqnarray}
and hence, the net effect of a Kerr-Schild transformation in the
metric functions is given by
\begin{eqnarray}
h_0(r)\to {h}(r)=h_0(r)-\mu a(r),\qquad
f_0(r)\to{f}(r)=\frac{f_0(r)\left(h_0(r)-\mu a(r)\right)}{h_0(r)}.
\label{KStransf}
\end{eqnarray}
Most importantly we need to specify the change of the scalar field through the Kerr-Schild transformation. In analogy with the
Kerr-Newman example (\ref{normA}), we will demand that the kinetic term of the scalar field denoted by $X$ remains unchanged
({\it{but not necessarily constant}}) under the Kerr-Schild transformations of the metric, i. e.
\begin{eqnarray}
X^{(0)}:=g^{(0)\mu\nu}\,\partial_{\mu}\phi^{(0)}\,\partial_{\nu}\phi^{(0)}=X:=g^{\mu\nu}\,\partial_{\mu}\phi\,\partial_{\nu}\phi.
\label{kineticInv}
\end{eqnarray}
At a first glance, this condition seems to be quite restrictive and
not natural. Indeed, it is simple to see that this condition does
not hold for the black hole solutions of scalar fields nonminimally
coupled to Einstein gravity \cite{Bekenstein:1974sf,
Martinez:2002ru}. Nevertheless, as shown below, for general DHOST
theories with shift and parity symmetries,
$\phi\to\phi+\mbox{constant}$ and $\phi\to -\phi$, the set of
equations is always invariant under the condition
(\ref{kineticInv}).

Let us specify concretely the procedure to follow. Our starting
point is the static seed configuration (\ref{seedconf}) which is
solution of some DHOST theory described by the action
$$
S(g,\phi)=\int d^4x\sqrt{-g}\, {\cal L}_{\mbox{{\tiny DHOST}}}.
$$
We will say that this theory is {\it Kerr-Schild invariant} with the
transformations (\ref{KStransf}-\ref{kineticInv}) if the variation
of the action is quasi-invariant\footnote{By quasi-invariant, we
mean an action invariant up to boundary terms denoted by
$\mbox{b.t.}$.} for a specific selection $a(r)=a(r, X(r))$.
Concretely, this will be achieved if
\begin{eqnarray}
S(g,\phi)-S(g^{(0)},\phi^{(0)})=\int dr\,{\cal E}\left(r, a(r),
a^{\prime}(r), X(r), X^{\prime}(r)\right)+ b.t., \label{varS}
\end{eqnarray}
for a function  $a(r)=a(r, X(r))$ solving the differential equation
given by ${\cal E}(r, a(r), a^{\prime}(r), X(r), X^{\prime}(r))=0$.
We stress that the possible dependence of the scalar function $a$ on
the kinetic term $X(r)$ can be justified from the fact that this
latter is the unique scalar quantity that remains invariant under
the Kerr-Schild procedure (\ref{kineticInv}). In practice, this can
be viewed as a generating method to easily construct black hole
configurations from a simple seed solution. Let us illustrate the
Kerr-Schild process with the simple example of the shift-symmetric
quadratic Horndeski theory,
\begin{eqnarray}
S[g, \phi]=\int d^4x\sqrt{-g}\,
\Big\{G_2(X)+G_4(X)R-2\,G_{4,X}(X)\left[\left(\Box\phi\right)^2-\left(\nabla_{\mu}\nabla_{\nu}\phi\right)
\left(\nabla^{\mu}\nabla^{\nu}\phi\right)\right]\Big\}.
\label{actionG2G4}
\end{eqnarray}
After some manipulations, the variation of this action  through the
Kerr-Schild transformations (\ref{KStransf}-\ref{kineticInv}) yields
(up to a boundary term)
\begin{eqnarray}
S(g,\phi)-S(g^{(0)},\phi^{(0)})=-8\pi\mu\int
dr\sqrt{\frac{f_0(r)}{h_0(r)}}\,\Big[a(r)+r
a^{\prime}(r)\Big]\Big[-2G_{4,X} X+G_4\Big].
\end{eqnarray}
Hence, we conclude that starting from any seed configuration
(\ref{seedconf}) solution of the quadratic Horndeski theory
(\ref{actionG2G4}), the transformed metric and scalar field defined
respectively by (\ref{metrKS}) and (\ref{kineticInv}) will be a
black hole solution of the same theory provided that the metric
function $a$ satisfies $a(r)+ra^{\prime}(r)=0$, in other words for a
Coulomb fall off, $a(r)=\frac{1}{r}$. It is also interesting to
stress that this Coulomb behavior for the mass term is independent
of the seed configuration. This is in complete accordance with the
existence of black hole solutions for this quadratic Horndeski
theory which are asymptotically flat and (A)dS, see Refs. \cite{
Babichev:2013cya, Kobayashi:2014eva, Kobayashi:2011nu,
Anabalon:2013oea, Lehebel:2018zga, Babichev:2017guv} but also for
solutions with rather exotic asymptotics,
\cite{Bravo-Gaete:2013dca}.

What is more interesting with this generating method is that we can
go further than GR and construct exotic solutions such as regular
black holes in scalar tensor theories (see for example \cite{Bronnikov:2018vbs}). Such solutions were thought
to exist in higher order theories (see for example chapter 8 of \cite{Chinaglia:2018uol}) but the vast functional space of the
coupling functions made the theories and solutions hard to find. We
will achieve this here, precisely due to the Kerr-Schild ansatz, as
the seed solution and the mass function $a$ can be found in two
{\it{independent steps}}.  We will see that we can first adequately
fix the mass function and then reverse engineer the seed theory in
order to find a full regular black hole solution. It is very
probable that our method can be extended to different asymptotic
spacetimes as well as other exotic solutions such as wormholes or
solitons.

The paper is organized as follows : in section $2$ we shall
acquire the validity of the Kerr-Schild ansatz in general
higher order scalar-tensor theories. In particular, we will see that certain
degeneracy conditions, necessary for healthy theories, will actually
be required for the validity of Kerr-Schild symmetries while on the
other hand,  certain healthy scalar tensor theories such as the DGP
Galileon, or the Gauss-Bonnet, will not adhere to the Kerr-Schild
symmetries. We will then find new static black hole solutions, with
differing mass fall-off, with relative ease within very general DHOST
theories. With hindsight from section $2$ we will then move on to
construct black holes without a central curvature singularity. As a
direct application of this procedure, we will present a special
DHOST theory that sources the regular Hayward metric black hole
\cite{Hayward:2005gi}. In the last section, we will summarize our
results and discuss some generalizations of our work (in particular
for generalized Proca theories). Finally, an appendix will be
devoted to the three-dimensional rotating case of the quadratic
Horndeski theory.
%
%

\section{ Kerr-Schild symmetry for generalized  scalar tensor theories}

Before treating the physical case of DHOST theories, we first consider a more  general class of scalar tensor
theories which contains up to second order covariant derivatives of the scalar field and
whose action is,
\begin{eqnarray}
S[g,\phi]=\int
d^4x\sqrt{-g}\Big[&K(X)+G(X)R+F_1(X)\Box\phi+F_2(X)G^{\mu\nu}\phi_{\mu\nu}+A_1(X)\phi_{\mu\nu}\phi^{\mu\nu}+A_2(X)(\Box\phi)^2
\nonumber\\
&+A_3(X)\Box\phi\,\phi^{\mu}\phi_{\mu\nu}\phi^{\nu}+A_4(X)\phi^{\mu}\phi_{\mu\nu}\phi^{\nu\rho}\phi_{\rho}+A_5(X)\left(\phi^{\mu}\phi_{\mu\nu}\phi^{\nu}\right)^2\Big].
\label{action}
\end{eqnarray}
This theory includes shift-symmetric Horndeski and DHOST theories
but also theories which can have Ostrogradski ghosts. Our aim is to
check when the Kerr-Schild ansatz is valid and if certain conditions
of degeneracy are required by the ansatz itself. For simplicity we
have defined $\phi_{\mu\nu}=\nabla_{\mu}\nabla_{\nu}\phi$ and $K, G,
F_1, F_2$ and the $A_i$ are a priori arbitrary functions of the
kinetic term $X=\partial_{\mu}\phi\,\partial^{\mu}\phi$ which in
turn ensures the shift invariance  $\phi\to\phi+\mbox{cst}$ of the
action. For latter convenience, we also define
\begin{eqnarray}
&&\HH(X)=A_1(X)\,X-G(X),\qquad \B(X)=A_3(X)\,X+4G_X(X)-2A_1(X),\nonumber\\
&&\ZZ(X)=A_3(X)+A_4(X)+X\,A_5(X).
\label{defHBZ}
\end{eqnarray}
As previously, we are interested under which conditions the action
(\ref{action}) is (quasi)invariant under a Kerr-Schild
transformation (\ref{KStransf}) leaving invariant the kinetic term
(\ref{kineticInv}). It is interesting to note that a necessary
condition for the action to be (quasi)invariant is to impose
$A_2(X)=-A_1(X)$ and $F_1=F_2=0$. While the former is the Horndeski
degeneracy condition the latter conditions tell us that theories
without parity symmetry ($\phi\leftrightarrow -\phi$) are excluded
from the Kerr-Schild solution generating method. This includes
therefore the $G_3$ and $G_5$ Galileons (see for example
\cite{Kobayashi:2011nu}). Only under these conditions, the variation
of the action can be of the form of Eq. (\ref{varS}), i. e.
\begin{eqnarray}
\label{var} S[\bar{g},\bar \phi]-S[g,\phi]=-\mu \pi \int
dr\sqrt{\frac{f_0(r)}{h_0(r)}}\Big[a(r) P(r,X)+a^{\prime}(r)
Q(r,X)\Big]+ b.t.,
\end{eqnarray}
where the expressions of $P$ and $Q$ read
\begin{eqnarray}
P(r,X)=r^2(X^{\prime})^2\ZZ(X)+4rX^{\prime}\B(X)-8\HH(X),\qquad Q(r,X)=r^2X^{\prime}\B(X)-8r\ZZ(X).
\end{eqnarray}
Hence, in order for the transformations (\ref{KStransf}-\ref{kineticInv}) to be a
symmetry of the action, the function $a$ must be given by
\begin{eqnarray}
a(r)=e^{-\int \frac{P(r,X)}{Q(r,X)}dr}, \label{sol}
\end{eqnarray}
provided $Q(r,X)\not=0$, and where of course the integration
constant can be re-absorbed into the mass parameter $\mu$. Note that
$K(X)$ is not involved in the above mass integral in accord with our
requirement for $X$ (\ref{kineticInv}). Also, in accordance to
classical no hair theorems \cite{Bekenstein} the vanilla case of a
minimally coupled scalar field ($K=-\frac{1}{2}X$, $G=1$), i. e.
$\HH(X)=-1$ and $\B(X)=0$,  is clearly excluded. It is also
interesting to note from the equation (\ref{sol}) that the standard
Coulomb fall off for the mass $a(r)\propto \frac{1}{r}$ will be
ensured only if $Q(r,X)=r P(r,X)$. This condition is achieved for
constant kinetic term $X=\mbox{constant}$ or for
\begin{eqnarray}
r\ZZ(X)\, X^{\prime}+3\B(X)=0.
\label{constCoulomb}
\end{eqnarray}
In particular, this last condition is satisfied for the quadratic
Horndeski case (\ref{actionG2G4}) in accordance with the results
presented in the introduction. Going beyond the quadratic Horndeski
case therefore already hints on a much richer behavior with
differing fall-off mass functions.

\subsection{Degenerate Higher Order Scalar Tensor Theories (DHOST)}
We now move on to consider the physical subclass of the scalar tensor
theories (\ref{action}) invariant under the reflection $\phi\to -\phi$ and free
of  Ostrogradski ghosts \cite{Langlois:2015cwa, Langlois:2015skt, Crisostomi:2016czh}.  The first condition imposes that $F_1=F_2=0$ while the absence of Ostrogradski
ghosts may be ensured through the following restrictions,
\begin{eqnarray}
\label{2n1} A_1&=&-A_2\neq \frac{G}{X},
\nonumber\\
A_4&=&\frac{1}{8(G-XA_1)^2} \left\{ 4G \left[
3(-A_1+2G_{X})^2-2A_3G\right] -A_3X^2(16A_1G_{X}+A_3 G) \right.
\nonumber\\
&& \left. + 4X \left[
-3A_2A_3G+16A_1^2G_{X}-16A_1G_{X}^2-4A_1^3+2A_3G G_{X} \right]
\right\},
\nonumber\\
A_5&=&\frac{1}{8(G-X A_1)^2} (2A_1-XA_3-4G_{X})
\left(A_1(2A_1+3XA_3-4G_{X})-4A_3G\right).
\end{eqnarray}
Note that while $A_1=-A_2$ was also necessary for the validity of
the Kerr-Schild ansatz the latter two conditions are not.

\subsubsection{DHOST solutions with a time independent scalar field $q=0$}

For simplicity, let us consider  the case where the scalar field
does not depend on time, i. e. $q=0$. This case has been considered
recently in \cite{Minamitsuji:2019tet} and some solutions with $X$
non constant were given. Using the notations (\ref{defHBZ}), the field equations
can be written in a tractable way as
\begin{subequations}
\label{eqsfund}
\begin{eqnarray}
&&X[2(A_1 G)_X+G A_3]+r^2\left[(K\HH)_X+\frac{3}{4} K \B\right]=0,\label{seed1}\\
&&-3(\B r X')^2 f+8 (\B r X')f \HH \left(\frac{rh'}{h} +4\right) - 32 f\HH \left[\frac{K r^2 +2 G}{f}+2\HH \left(\frac{rh'}{h} +1\right)\right]=0,\label{seed2}\\
&& r^2(16 \B_X \HH + 3 \B^2) X'^2+8\HH X' r\left(\B r \frac{f'}{f}-16
\HH_X\right)+16 r^2 \HH \B X''-64 \HH^2\left[\left(\frac{rf'}{f} +1 \right) +\frac{2 G +
r^2 K}{2f\HH}\right]=0,\nonumber\\
\label{seed3a}
\end{eqnarray}
\end{subequations}
where we have assumed that $\HH\neq 0$ and $K r^2 +2 G \neq 0$. The
equations (\ref{seed1}) and (\ref{seed2}) are identical to those
given in \cite{Minamitsuji:2019tet} while the latter is the
$tt-$equation which simplifies considerably in this notation. It is
interesting to note that (\ref{seed1}) fixes $X$ independently of
the spacetime metric \cite{Minamitsuji:2019tet}, condition that is
in adequation with the invariance of the kinetic term
(\ref{kineticInv}) through a Kerr-Schild transformation.

Using these variables it is straightforward to verify that the mass
function for the Kerr-Schild transformation (\ref{sol}) is given by,
\be a(r)=\frac{1}{r} e^{\frac{3}{8}\int dX \; \frac{\B}{\HH}}.
\label{MMM} \ee The same conclusion can be obtained from the field
equations (\ref{eqsfund}). As already mentioned, for invariant $X$
as defined by (\ref{kineticInv}), the equation (\ref{seed1}) is
trivially invariant under the Kerr-Schild transformation of the
metric (\ref{KStransf}) while the invariance of the Eq.
(\ref{seed2}) will be ensured only if $a(r)$ is given by
(\ref{MMM}). Finally, the remaining equation (\ref{seed3a}) under
the Kerr-Schild transformations (\ref{kineticInv}-\ref{KStransf})
with $a$ given by (\ref{MMM}) will be mapped to itself provided Eq.
(\ref{seed2}) is satisfied.

Rather than solving the curved field equations given above it
suffices to find a seed solution and use the Kerr-Schild
transformation to evaluate the mass function (\ref{MMM}). We will
choose to do this in the case of flat spacetime although numerous
other cases are possible such as dS, adS, or other more exotic
vacua. One needs only to put the relevant metric ansatz for the seed
metric solution in order to proceed. For a flat seed metric
$f_0=h_0=1$, the field equations read,
\begin{subequations}
\label{eqfh1}
\begin{eqnarray}
&&X[2(A_1 G)_X+G A_3]+r^2\left[(K\HH)_X+\frac{3}{4} K \B\right]=0,\label{seed1a}\\
&&-3(\B r X')^2+32 (\B r X') \HH - 32 \HH (K r^2 +2 A_1 X)=0,\label{seed2a}\\
&&r^2(16 \B_X H + 3 \B^2) X'^2-128 r \HH \HH_X X'+16 r^2 \HH \B X''-32
\HH (2 A_1 X + r^2 K)=0.\label{seed3}
\end{eqnarray}
\end{subequations}
Given a particular theory, equation (\ref{seed1a}) will provide an
algebraic expression for $X$ whereas the remaining equations
(\ref{seed2a}) and (\ref{seed3}) will impose compatibility
constraints onto the coupling constants and parameters of the model
under consideration. In fact, from (\ref{seed2a}) we obtain \be \B
rX'=\frac{16 \HH}{3}\left(1 \pm \sqrt{1-\frac{3 (K r^2+ 2 A_1 X)}{8
\HH}}\right)  \label{cc1} \ee with $\HH\neq 0$, whereas from
(\ref{seed3}) and (\ref{seed2a}) we have \be \label{cc2}
 -(\B r^2 X')'+ 8 r \HH_X X'+ 4 (2 A_1 X+ r^2K)=0
\ee
Due to the structure of the
field equations, one approach is to keep $\B$ and $\HH$ of a given
order, hence we consider
\begin{equation}
\label{ansatz}
G=\gamma X^n, \hspace{0.2cm}A_1=\alpha X^{n-1}, \hspace{0.2cm}
A_3=\beta X^{n-2}, \hspace{0.2cm} K=-\Lambda X^m
\end{equation}
This choice directly allows us to evaluate the mass term,
\begin{align}
\HH&=(\alpha-\gamma)X^n=\HH_0 X^n\\
\B&=(\beta+4 \gamma n-2 \alpha)X^{n-1}= \B_0 X^{n-1}
\end{align}
and therefore \be a(r)=\frac{1}{r} X^{\frac{3\B_0}{8\HH_0}}
\label{mass0} \ee Second, from (\ref{seed1}) we can algebraically
solve the kinetic term $X(r)$ explicitly
\begin{equation}
X^{n-m}=\delta r^2   \label{solution}
\end{equation}
where \be \delta=\Lambda\frac{4 m \alpha-4 m \gamma + 3 \beta + 4
\alpha n - 6 \alpha + 8 \gamma n}{4 \gamma(-2 \alpha + \beta + 4
\alpha n )} =\Lambda \frac{4(m+n)\HH_0+3\B_0}{4\gamma (\B_0+4 n
\HH_0)}.\label{del} \ee
Finally, we note that (\ref{seed2}) and (\ref{seed3})  reduce into
two constraints for the theory, \be
\B_0=(n-m)\frac{8\HH_0}{3}\left(1\pm \sqrt{1-3 \frac{2 \delta \alpha
- \Lambda}{8 \HH_0 \delta}} \right)   \label{cons1} \ee \be \B_0
\frac{3n-m}{n-m}= 8 \HH_0 n + \frac{2(2 \alpha \delta-
\Lambda)(n-m)}{\delta}   \label{cons2} \ee
Straightforwardly, equations (\ref{cons1}) and (\ref{cons2}) can be
re-written in the following manner
\begin{equation}
\begin{split}
\label{c02}
\frac{3\B_0}{8\HH_0}&=(n-m)(1\pm\sqrt{1-\epsilon}),\\
\frac{3\B_0}{8\HH_0}&=\frac{n-m}{3n-m}(3n+2(n-m)\epsilon),
\end{split}
\end{equation}
where we have defined the auxiliary quantity,
$\epsilon=\frac{3}{8\HH_0}\frac{2\delta\alpha-\Lambda}{\delta}$,
which is required to be smaller than one. This form of the
constraints gives direct access to the form of the mass function
(\ref{mass0}). On the other hand, compatibility of (\ref{c02})
implies that
\begin{equation}
1\pm\sqrt{1-\epsilon}=\frac{3n}{3n-m}+2\epsilon\frac{n-m}{3n-m}
\end{equation}
must be satisfied. The later condition defines two branches of
solutions:
\subsubsection*{$\bullet$ Branch 1}
For this branch we have that
\begin{equation}
1+\sqrt{1-\epsilon}=\frac{3n}{3n-m}+2\epsilon\frac{n-m}{3n-m}.
\end{equation}
Solving for $\epsilon$ implies $\epsilon=3/4$ which gives the following constraint on the parameters,
\begin{equation}
\label{br1cond}
4m\alpha-2\alpha+\beta-4m\gamma-4n\alpha+8\gamma n=0.
\end{equation}
It so happens that the above condition satisfies both equations
(\ref{c02}). By plugging this result into the mass function
(\ref{mass0}) we observe that there is no dependence on the
parameters, giving a mass function of the form
\begin{equation}
\label{br1a}
a(r)=r^2
\end{equation}
where in this case the integration constant $\mu$ as defined through
$\mu a(r)$ in the metric  (\ref{metrKS}) plays the role of a
cosmological constant changing the asymptotic behavior of the
solution relative to the size and magnitude of $\mu$.

However, the fact that both of the equations~(\ref{c02}) are satisfied with one condition~(\ref{br1cond})
signals degeneracy of the equations. This happens because the choice~(\ref{ansatz}) with (\ref{br1cond}) gives $2G+ r^2 K=0$ for which equations (\ref{seed1a}-\ref{seed3}) are not satisfied.
At the end a careful resolution of the initial Euler-Lagrange equations give us an additional condition,
\begin{equation}
\label{br1cond2}
\alpha\left(m-3n\right)+2n\gamma=0.
\end{equation}
To summarize, when both conditions~(\ref{br1cond}) and
(\ref{br1cond2}) are satisfied, Eq.~(\ref{ansatz}) and (\ref{br1a})
is solution. This solution is seemingly quite special as the 'mass'
integration constant behaves as a cosmological constant completely
independently of the bulk cosmological constant term obtained in $K$
(for $m=0$ for example).

\subsubsection*{$\bullet$ Branch 2}
For the negative branch we get
\begin{equation}
1-\sqrt{1-\epsilon}=\frac{3n}{3n-m}+2\epsilon\frac{n-m}{3n-m}
\end{equation}
As in the previous case we solve for $\epsilon$, which now reads
\begin{equation}
\epsilon=\frac{n(2m-3n)}{(m-n)^2}\leq 1 .
\end{equation}
Using the above relation, from (\ref{c02}) we obtain,
\begin{equation}
\frac{3\B_0}{8\HH_0}=-n.
\end{equation}
Solving the above two constraints gives us two relations between the coefficients of the Lagrangian,
\begin{equation}
\begin{split}
2n\gamma &= \alpha(3n-m),\\
3\beta &= 2(3 + m -7n )\alpha.
\end{split}
\end{equation}
Thus for this branch the mass function depends on $n$ and $m$,
\begin{equation}
a(r)=r^{-\frac{3n-m}{n-m}}.
\end{equation}
with the integration constant is given by $\mu$ in our
representation (\ref{metrKS}). This branch offers a far richer
structure for the solutions. We note that we get a Coulomb mass term
if $n=0$. The resulting metric after the Kerr-Schild procedure is
given by, $$f=h=1-\frac{\mu}{r^{\frac{3n-m}{n-m}}}$$ and describes an
asymptotically flat black hole spacetime for $m>3n$ and $m>n$. Note
that $X$ is then singular at $r=0$ while it goes to zero as $r\to
\infty$. The family of black hole solutions includes the case where we have a canonical kinetic term $m=1$. We also note the generality of the solution obtained with
relative ease, as integrating the equations with Kerr-Schild reduces
to finding a flat spacetime solution. Note for completeness that the
solution given here, agrees with the static asymptotically flat
solution found in \cite{Minamitsuji:2019tet}, by solving the curved field equations (\ref{seed1}-\ref{seed3a}) with
$m=\alpha=0$ upon which we have $a(r)=r^{-3}$.

\section{Regular black hole solutions}

In the previous section we constructed a large class of black hole
solutions in DHOST theories. Starting from a seed configuration,
represented by $(h_0, f_0, X_0)$ of some DHOST theory, we constructed a black hole solution with the same scalar kinetic term
$X=X_0$ (where $X=X_0$ is not necessarily constant) and a mass term
given by (\ref{sol}). There is an important technical advantage within the method, which distinguishes two independent steps : finding the seed solution and seeking the mass fall-off term.
We now apply this technique in order to construct a regular
black hole solution within DHOST theory. By regular black hole we mean a black hole solution with at least one
regular event horizon and no central curvature singularity.
The first such regular black hole geometries were
proposed in \cite{Bardeen} but no source of this solution was given.
Later on, the authors of Ref. \cite{AyonBeato:1998ub} were the first
to present an exact regular black hole solution of General
Relativity coupled to a specific nonlinear electrodynamic source.

From our Kerr-Schild procedure, looking for an asymptotically flat
regular black hole with a seed metric given by $h_0=f_0=1$, will
imply that the final metric will be of the form (\ref{sol})
$$
h(r)=f(r)=1-\frac{\mu}{r}e^{\int dX\,\frac{3\B}{8\HH}}.
$$
Given the form of the above mass function we can make the following simple hypothesis,
\begin{equation}
\frac{3\B}{8\HH}=\frac{\lambda}{X} \Longrightarrow
h(r)=f(r)=1-\frac{\mu \;X(r)^\lambda}{r}, \label{hyp1}
\end{equation}
where $\lambda$ is a constant. The idea will be to choose an
appropriate kinetic term $X(r)$ and a parameter $\lambda$ in order
for the metric to be : regular at $r=0$, asymptotically flat and to
have an outer event horizon at some finite $r=r_h$. This would mean that
our seed configuration is given by $h_0=f_0=1$ together with $X(r)$,
and hence by inverse engineering from the field equations
(\ref{eqfh1}), one would be able to specify the corresponding DHOST
theory, that is to determine the functions $K, G, A_1$ and $A_3$ (as
functions of $X$ only). Keeping this in mind we start with
hypothesis (\ref{hyp1}) and solve for the underlying theory
functions, $K, G, A_3,...$ etc in terms of $X$, its derivatives and
$r$. It is important to stress that once we fix $X=X(r)$, at the end
we must have that all coupling functions are solely $X-$dependent
and not $r$-dependent. This will be achieved if the function $X$ is
locally invertible.

Let us proceed by detailing all the different steps of the
construction. We first observe that using the hypothesis
(\ref{hyp1}), the constraints (\ref{cc1}) and (\ref{cc2}) take the
form
\begin{align}
\lambda \frac{rX^{\prime}}{X}&=2(1\pm\sqrt{1-\epsilon})\label{cn1}\\
\left[r\HH\left(\lambda\frac{rX^{\prime}}{3X}-1\right)\right]^{\prime}&=\left(\frac{4}{3}\epsilon-1\right)\HH,
\label{cn2}
\end{align}
where for simplicity we have defined
\begin{equation}
\epsilon=\frac{3(Kr^2+2A_1X)}{8\HH}.
\end{equation}
In addition, from Eq. (\ref{hyp1}) and making use of the definition
of $\B$, we can trade off $A_3$ and $A_1$
\begin{align}
A_3=-\frac{4G_{X}}{X}+\frac{2A_{1}}{X}+\frac{8\lambda\HH}{3X^2}\qquad
A_1=\frac{\HH+G}{X} \label{id1}
\end{align}
and, hence through the  definition of $\epsilon$ one can get either
$G$ or $K$ as
\begin{equation}
\label{id2} G+\frac{1}{2}K r^2= \left(\frac{4
\epsilon}{3}-1\right)\HH.
\end{equation}
After some algebraic manipulations, Eq. (\ref{seed1}) is
conveniently reduced as
\begin{equation}
2(\HH
G)_X+r^2(K\HH)_X+\frac{2\lambda\HH}{X}\left(\frac{4}{3}G+Kr^2\right)=0.
\label{edo}
\end{equation}
Now the picture of the construction is quite clear. From Eq.
(\ref{cn1}), the $\epsilon-$ function is obtained, and this allows
us to get $\HH$ by solving Eq. (\ref{cn2}). Finally, using the last
equation (\ref{edo}), we will get $K$. In practice, we end up with
\begin{eqnarray}
\epsilon=\left( 2-\frac{\lambda r X'}{2 X}\right)\frac{\lambda r
X'}{2 X},\qquad \HH=\frac{H_0}{X^\lambda \left( \frac{\lambda r
X'}{3 X}-1\right)},\qquad K = -\frac{\A_{,r}}{r X^{\lambda/3}},
\end{eqnarray}
where $H_0$ is an integration constant, and where $$\A=\frac{H_0
\left( 1-\frac{\lambda r X'}{ X}\right)}{ X^{2\lambda/3}\left(
\frac{\lambda r X'}{3 X}-1\right)}.$$ Having all the functions fixed
in terms of $r$ and $X(r)$, we will now implement our second
hypothesis. We choose a kinetic term $X(r)$ that is an invertible
function in a such way that combined with a judicious choice of
$\lambda$, it will provide a regular black hole solution. In order
to ensure the regularity of the solution, the metric should satisfy
the following requirements: $(i)$ it must be regular at the origin
$r=0$ and at infinity (asymptotically flat in our case), $(ii)$ it
must possess an outer event horizon at some finite $r=r_h$, and
finally $(iii)$ the metric function $f=h$ must satisfy the Sakharov
criterion \cite{Sakharov:1966aja, Chinaglia:2018uol} at the origin,
i. e.
\begin{equation}
f(r)\displaystyle\mathop{\sim}_{r\sim 0} 1- f_0 r^p,\qquad p\geq2,
\end{equation}
which means that the metric function possesses {\it at least} a
de-Sitter core near the origin. The condition $(iii)$ ensures that
any invariant constructed from the contractions of the Riemann curvature tensor will be regular at the origin. \\
We observe that a relatively simple and invertible function $X(r)$
that provides a metric function satisfying the conditions $(i)$ and
$(ii)$ is given by
\begin{equation}
X(r)=\frac{\alpha r^2}{\beta r^2+\gamma},\Longrightarrow
r^2=\frac{\gamma X}{\alpha-\beta X} \label{inverse}
\end{equation}
where $\alpha, \beta$ and $\gamma$ are arbitrary non-zero real
constants independent of the mass, while the choice of $\lambda=2$ ensures that condition
$(iii)$ is satisfied{\footnote{Note that one of the three $\alpha, \beta$ or $\gamma$ can be set to 1 without loss of generality.}}. Finally, using the relation (\ref{inverse}) we
easily conclude that the  DHOST theory parameterized by
\begin{eqnarray}
&& {\cal H}(X)=\frac{H_0 \a^{3}}{X^2(4 \b X- \a)},\qquad G(X)=-\frac{3H_0\a^2\left(8\b^2X^2-8\a\b X+\a^2\right)}{X^2(4 \b X- \a)^2},\nonumber\\
&& K(X)=\frac{8\a
H_0\left(16X^4\b^4-54X^3\a\b^3+63X^2\b^2\a^2-28\a^3\b
X+3\a^4\right)}{X^3\gamma\left(\a-4\b X\right)^2},
\end{eqnarray}
supports the existence of the following spherically symmetric
regular black hole
\begin{equation}
ds^2=-\left(1-\frac{ \mu \a^2 r^3}{(\b
r^2+\gamma)^2}\right)dt^2+\frac{dr^2}{\left(1-\frac{\mu \a^2
r^3}{(\b r^2+\gamma)^2}\right)}+r^2(d\theta^2+\sin^2\theta
d\varphi^2) \label{regular}
\end{equation}
It is easy to observe that solution (\ref{regular}) is well behaved
at the origin and at infinity, with a Kretchmann invariant
\begin{equation}
\lim_{r\rightarrow 0}R_{\mu\nu\rho\sigma}R^{\mu\nu\rho\sigma}\sim
76\left(\frac{\a}{\gamma}\right)^4r^2 + \mathcal{O}(r^4)
\end{equation}
along with a regular everywhere kinetic term $X(r)$. Furthermore, for $\alpha>0$ large enough (compared to $\beta$)
the metric will have an inner and outer event horizon at some positive values of the radial coordinate $r$.
One can of course construct, in a similar fashion, different regular metrics.
To end this section, let us instead
construct the specific DHOST theory that sources a known regular geometry, the Hayward metric, \cite{Hayward:2005gi}. The interests of this spacetime example
are multiple and particularly used to test the formation and evaporation
of non-singular black holes, see e. g.  \cite{Frolov:2016pav}. For
our static ansatz, the  regular Hayward black hole metric is given
by
\begin{equation}
ds^2=-\left(1-\frac{\mu \a r^2}{\b
r^3+\gamma}\right)dt^2+\frac{dr^2}{\left(1-\frac{\mu \a r^2}{\b
r^3+\gamma}\right)}+r^2(d\theta^2+\sin^2\theta d\phi^2).
\label{Haywardmetric}
\end{equation}
From our previous derivation, this would correspond to $\lambda=1$
and
\begin{equation}
X(r)=\frac{\a r^3}{\b r^3+\gamma},
\end{equation}
and the following DHOST theory
\begin{eqnarray}
&& {\cal H}(X)=\frac{H_0 \a^{2}}{X^2},\qquad G(X)=\frac{H_0 \a(5\a-6\b X)}{X^2},\nonumber\\
&& K(X)=\frac{2H_0(\a-\b X)^2(3\b X-5\a)}{\gamma
X^3}\left(\frac{\gamma X}{\a-\b X}\right)^{\frac{1}{3}},
\end{eqnarray}
will source the regular metric (\ref{Haywardmetric}). Note that the
scalar field is non-trivial even when the mass parameter $\mu$ is
switched off. This may imply that in the ground state of the theory
the scalar is not constant, but rather it forms a soliton which does
not backreact on the metric.

\section{Conclusions}
In this paper, we have explored an extension of the Kerr-Schild
solution generating method for shift invariant higher order scalar
tensor theories. In order to achieve this, we have put forward a
Kerr-Schild ansatz for a static configuration including the metric
(similarly to GR) and the kinetic term of the scalar field. We have
found that quite generically DHOST theories with parity invariance
(for the scalar field) obey the Kerr-Schild symmetries with some
mass function, that is Coulombian for Horndeski, and possibly
non-Coulombian once we go to DHOST theories. It is well known that
in GR, the Kerr-Schild method reproduces most known interesting
solutions. It is hardly ever the method that is initially used to
obtain some solution. This is actually true for most solution
generating methods in GR. On the contrary here, for scalar tensor
theories, given the complexity of the field equations, and the
generality of the coupling functions at hand, we have shown that the
method is actually crucial in obtaining new solutions such as black
holes or even more exotic vacua like regular black holes. What is
important here is the independent two step obtention of the seed
metric and the mass fall-off function. It is very probable  that our
method can be extended to other seed vacua as de-Sitter or even
other exotic solutions such as wormholes or solitons. In this sense
it is possible that the Kerr-Schild extension may help significantly
in obtaining new solutions in scalar tensor theories.

{When studying modified gravity theories, it is important to search for novel solutions and
to compare them with the solutions in GR.
In the case of DHOST theories, for most of the solutions presented in the literature (with a few exceptions)
the metric is the same as in GR, such as Schwarzschild-dS or Kerr-dS metrics.
In this paper we presented a method for constructing solutions with metric other than known GR metric.
Comparing properties of new solutions with their counterparts in GR will help to search for signatures of modified gravity by using modern and future gravity experiments, and, on the other hand, to place constraints on particular  modifications of gravity.
In particular, the trajectories of stars around supermassive black holes would be clearly modified, giving an opportunity to test the theory.
In addition, it would be also interesting to study physical properties of the obtained solutions as compared to GR case.
One may expect that the thermodynamics of black holes is considerably modified for the solutions presented in the paper.}



Apart from these general observations a natural extension to our
work is generating a procedure for stationary and axisymmetric
configurations. This task is far from obvious but a first step in
this direction has been made in the three-dimensional quadratic
Horndeski theory with a scalar field depending linearly on the time
and angular coordinates (see the Appendix). Here, the challenges to
be met in 4 dimensions are numerous as for example the fact that
stationary vacua do not have the circularity property (unlike Ricci
flat or Einstein spaces). This already renders difficult the
starting ansatz for the seed metric. For certain stealth solutions
as the ones reported in \cite{Ruth} our method may be generalized on
very similar lines as presented here.

Even if the working examples were for purely radial scalar field
$\phi=\phi(r)$, our procedure for constructing solutions also
applies for a scalar field depending linearly and separately on the
time coordinate, i. e. $\phi(r,t)=qt+\psi(r)$. It will be of
interest to look for such configurations in spite of the fact that
the presence of a time dependent scalar fields considerably complicates
the field equations even for a simple seed configuration. Note that
such a solution has been found in Ref. \cite{Minamitsuji:2019tet} (see
Eqs. (66)-(68-70)) for a subclass of the DHOST theory parameterized
by $\B=\frac{\gamma}{4}$ and $ \HH=\frac{\gamma \zeta}{X}$, and with
a kinetic scalar field given by
$$
X(r)=-q^2
\left(1-\frac{3r^2\Lambda}{4\zeta}\right)^{-\frac{4}{4-3\gamma}}.
$$
It is interesting to observe that a direct application of our mass
term (\ref{sol}) yields
$$
a(r)\propto
\frac{1}{r}\left(1-\frac{3r^2\Lambda}{4\xi}\right)^{\frac{3\gamma}{8-6\gamma}},
$$
and perfectly fits with the mass term of the solution reported in
Ref. \cite{Minamitsuji:2019tet}.

It will be desirable to extend the solution generating method for other
matter sources. As a natural candidate, and in a complete analogy
with the quadratic Horndeski case, we may consider the generalized
Proca action
\begin{eqnarray}
S_{GP}=\int d^4x\sqrt{-g}\Big[G_2(X, F)+
G_4(X)R+G_{4,X}\left((\nabla_{\mu}A^{\mu})^2-\nabla_{\mu}A_{\nu}\nabla^{\nu}A^{\mu}\right)\Big],
\label{geneProca}
\end{eqnarray}
where $X=-A_{\mu}A^{\mu}/2$ and $F$ is the standard Maxwell term
$F=-F_{\mu\nu}F^{\mu\nu}$. The question is to see under which
conditions the generalized Proca action (\ref{geneProca}) can be
invariant under the Kerr-Schild transformations. These
transformations will be the same for the metric (\ref{KStransf})
while the scalar field condition (\ref{kineticInv}) can be naturally
replaced by
\begin{eqnarray}
X^{(0)}:=-\frac{1}{2}\,g^{(0)\mu\nu}\,A^{(0)}_{\mu}A^{(0)}_{\nu}=X:=-\frac{1}{2}\,g^{\mu\nu}\,A^{}_{\mu}A^{}_{\nu},\qquad
F^{(0)}=F.
\end{eqnarray}
Nevertheless, it is simple to observe that in contrast with the
scalar field case, the on-shell action $S_{GP}$ on a purely electric
ansatz will explicitly depend on the electric potential $A_t(r)$ and
and its derivative. As a direct consequence, in order to properly
define the notion of Kerr-Schild invariance, this will require that
the electric potential $A_t(r)$ does not vary under
the Kerr-Schild transformations. On the other hand, since the Proca
mass term must remain constant, we are forced to turn on other
components of the potential vector $A_{\mu}$. The most simple option
in accordance with some solutions present in the literature \cite{Chagoya:2016aar}
is to consider a non-zero radial component, and hence
consider an ansatz configuration of the form
\begin{eqnarray}
ds^2=-h(r)dt^2+\frac{dr^2}{f(r)}+r^2\left(d\theta^2+\sin^2\theta
d\varphi^2\right),\qquad A_{\mu}dx^{\mu}=A_t(r)dt+A_r(r)dr.
\label{ansatzProca}
\end{eqnarray}
As anticipated previously, the on-shell action on the ansatz
(\ref{ansatzProca}) depends explicitly on $A_t(r)$, and reads
\begin{eqnarray}
S_{GP}=&&\int dr\,r^2\sqrt{\frac{h}{f}} G_2\left(X, \frac{f
(A_t^{\prime})^2}{2h}\right)\\
&&-\int dr
\sqrt{\frac{h}{f}}\Big\{G_4(X)\left[rf^{\prime}+f-1\right]+G_{4,X}(X)\left[-r(A_t^2)^{\prime}+2hrX^{\prime}-A_t^2+2rh^{\prime}X+2hX\right]\Big\}.\nonumber
\end{eqnarray}
It is interesting to note that the terms involving the electric
potential $A_t$ are expressions that remain invariant under the
Kerr-Schild transformations (\ref{KStransf}) with $A_t$ and $X$
being unchanged. Consequently, the variation under the Kerr-Schild
transformations of the generalized Proca action does not involve the
electric potential, and is given by (up to a boundary term)
\begin{eqnarray}
S_{GP}(\bar{g},A_t, X)-S_{GP}({g},A_t, X)=-2\mu\int
dr\,\sqrt{\frac{f}{h}}\left[2G_{4,X}X-G_4\right]\left[a+ra^{\prime}\right].
\end{eqnarray}
Finally as in the quartic Horndeski case, the mass term will be of
the Coulombian form, i. e.  $a(r)=\frac{1}{r}$ in accordance with
the existing solutions \cite{Chagoya:2016aar, Heisenberg:2017hwb}.
It will be also interesting to investigate if such invariance can be
exported to more general vector tensor theories
\cite{Heisenberg:2017hwb}.

\section*{Acknowledgments}
We would like to thank Eloy Ay\'on-Beato, Olaf Baake and Antoine
Leh\'ebel for many enlightening  discussions.
 A. C. and M. H. would like to thank the LPT/IJCLab of Orsay for
their kind hospitality during the elaboration of this project. E. B.
and C. C. acknowledge support from the CNRS grant 80PRIME and the
grant CNRS/RFBR Cooperation program for 2018-2020 n. 1985 Modified
gravity and black holes: consistent models and experimental
signatures.. A. C. work is supported by Fondo Nacional de
Desarrollo Cient\'ifico y Tecnol\'ogico Grant No. 11170274 and
Proyecto Interno Ucen I+D-2018, CIP2018020. The authors also
gratefully acknowledge the kind support of the PROGRAMA DE COOPERACI\'ON CIENT\'IFICA ECOSud-CONICYT 180011/C18U04.

\section*{Appendix: Example of rotating solution in three dimensions}

As a first approach to the stationary generalization of the
Kerr-Schild method let us present the three-dimensional case for the
quadratic Horndeski action,
\begin{eqnarray}
S[g,X]=\int
d^3x\sqrt{-g}\left\{G_2(X)+G_4(X)R-2G_{4,X}\left[\left(\Box\phi\right)^2-\phi_{\mu\nu}\phi^{\mu\nu}\right]\right\}.
\label{action2}
\end{eqnarray}
Assuming stationarity and since spacetime is circular, the most
general metric in $2+1$ dimensions can be parameterized as
\begin{eqnarray}
ds^2=-f(r)dt^2+\frac{dr^2}{f(r)}+H^2(r)\left[d\theta-k(r)dt\right]^2.
\label{ansatz2}
\end{eqnarray}
We will consider an ansatz for the scalar field,
\begin{eqnarray}
\phi(t,r,\theta)=qt +\psi(r)+L\theta,
\end{eqnarray}
where $q$ and $L$ are two constants. This ansatz for the scalar is
in accord with the Hamilton-Jacobi functional interpretation for
stationary spacetime geodesics \cite{Ruth}, and allows a mild linear
dependence in the $\theta$ and $t$ coordinates since
$\partial_\theta$, $\partial_t$ are two Killing vectors for
(\ref{ansatz2}). As always for the Kerr-Schild ansatz we first need
to find a seed flat solution of the above spacetime which
essentially sums up to finding the general solution given the
spacetime symmetry.

For the general $G_2$ and $G_4$ Horndeski theory,  the
$\epsilon_{tr}=0$ field equation permits to express the derivative
of the function $f$  as
\begin{eqnarray}
f^{\prime}=-\frac{1}{2}\frac{H\left[\left(2G_{4, X, X} X+G_{4,
X}\right)(k^{\prime})^2 H^2-2G_{2,X}\right]}{H^{\prime}\left(2G_{4,
X, X} X+G_{4, X}\right)}. \label{fprime}
\end{eqnarray}
Injecting this expression into the $\epsilon_{rr}=0$ equation, one
obtains that the kinetic term $X$ must satisfy the following
functional relation that only depends on $X$
\begin{eqnarray}
2 X\left[ G_{2,X} G_{4,X}+ G_2 G_{4, X, X}\right]-G_{2,X}G_4+G_2
G_{4,X}=0 \label{funrel}
\end{eqnarray}
which in turn implies that $X$ must be constant given by one of the
zeros of the previous functional equation. On the other hand, the
combination
$$
\frac{1}{2}\epsilon_{tt}+k(r)\epsilon_{t\theta}+\frac{k(r)^2}{2}\epsilon_{\theta\theta}=0,
$$
yields
\begin{eqnarray}
\frac{1}{2}f^2\left[2
G_{4,X}X-G_4\right]\frac{H^{\prime\prime}}{H}=0, \label{Hsecond}
\end{eqnarray}
and hence $H^{\prime\prime}=0$, and the last remaining Einstein
equation is equivalent to
\begin{eqnarray}
\Big(H^3 k^{\prime}\Big)^{\prime}=0. \label{kprime}
\end{eqnarray}
After some redefinitions of the radial and temporal coordinates, the
metric (\ref{ansatz2}) for the solution can be written as
\begin{eqnarray}
ds^2=-\frac{\beta}{4\alpha}\left(r^2-b^2\right)dt^2+\frac{dr^2}{\frac{\beta}{4\alpha}\left(r^2-b^2+\frac{\alpha
J^2}{\beta r^2}\right)}+J\,dt d\theta+r^2 d\theta^2, \label{btz}
\end{eqnarray}
where $b$ and $J$ are two integration constants, and where for
simplicity we have defined
\begin{eqnarray}
\alpha=2G_{4, X, X} X+G_{4, X},\qquad\beta=2G_{2, X}.
\end{eqnarray}
These expressions are evaluated on the constant value $X$ solution
of the equation (\ref{funrel}). Note that the metric solution is
nothing but the BTZ metric with a cosmological constant
$\Lambda=\frac{\beta}{4\alpha}$.

This three-dimensional case has been treated by brute force by
solving the field equations. It is then legitimate to wonder about
the validity of our Kerr-Schild mechanism in this three-dimensional
stationary case. The procedure can be formalized as follows.
Firstly, a null geodesic vector field for the seed metric
(\ref{ansatz2}) is given by $l=dt-\frac{dr}{f(r)}$, and hence the
Kerr-Schild transformation (\ref{kerrschild}) after the following
redefinitions of the angular and time coordinates as
$$
d\theta\to d\theta-\frac{\mu a(r) k(r)\,dr}{f(r)(f(r)-\mu
a(r))},\qquad dt\to dt-\frac{\mu a(r)\, dr}{f(r)(f(r)-\mu a(r))}
$$
brings the seed metric to the following same form
\begin{eqnarray}
ds^2=-\left(f(r)-\mu a(r)\right)dt^2+\frac{dr^2}{\left(f(r)-\mu
a(r)\right)}+H^2(r)\left[d\theta-k(r)dt\right]^2.
\label{kerrschildtrnasf}
\end{eqnarray}
In other words, the effect of the Kerr-Schild transformation on the
metric functions reads
\begin{eqnarray}
f(r)\to f(r)-\mu a(r),\qquad H(r)\to H(r),\qquad k(r)\to k(r).
\label{change3d}
\end{eqnarray}
Now, it is clear that the equations that fully characterize the
solution, namely  Eqs. (\ref{fprime}), (\ref{funrel}),
(\ref{Hsecond}) and (\ref{kprime}), are invariant under the
transformations (\ref{change3d}) only for a constant function $a$ as
we have obtained in the metric solution (\ref{btz}). The same
conclusion can be achieved from the variation of the action
(\ref{action2}) through the Kerr-Schild changes (\ref{change3d})
which yields
\begin{eqnarray}
\delta S=-2\pi\mu\int dr H^{\prime}(r) a^{\prime}(r)\left[-2G_{4,X}
X+G_4\right]+B. T.,
\end{eqnarray}
and hence one can again conclude that the mass term must be
constant.


\end{document}